\documentclass[8.5pt,twocolumn,oneside]{article}
\usepackage[super,sort&compress,comma]{natbib} 
\usepackage{times,mathptmx}
\usepackage{balance} 
\usepackage{graphicx} %eps figures can be used instead

\begin{document}

\twocolumn[
  \begin{@twocolumnfalse}
  
\center{\LARGE{\textbf{NMR Studies on the Temperature-Dependent Dynamics of Confined Water}}}

\vspace{0.6cm}

\large{\textbf{Matthias Sattig,\textit{$^{a}$}
Stefan Reutter,\textit{$^{a}$}
Franz Fujara,\textit{$^{a}$}
Mayke Werner,\textit{$^{b}$}
Gerd Buntkowsky,\textit{$^{b}$} and
Michael Vogel\textit{$^{a}$} 
}}\vspace{0.5cm}

\normalsize{
\textit{$^{a}$~Institut f\"ur Festk\"orperphysik, Technische Universit\"at Darmstadt, Hochschulstra\ss e 6, 64289 Darmstadt, Germany.}

\textit{$^{b}$~Eduard-Zintl-Institut f\"ur Anorganische und Physikalische Chemie, Technische Universit\"at Darmstadt,  Petersenstr. 20, 64287 Darmstadt, Germany. }

\vspace{0.5cm}

\textit{ E-mail: michael.vogel@physik.tu-darmstadt.de }
}
\vspace{1cm}

\normalsize{
We use $^2$H NMR to study the rotational motion of supercooled water in silica pores of various diameters, specifically, in the MCM-41 materials C10, C12, and C14. Combination of spin-lattice relaxation, line-shape, and stimulated-echo analyses allows us to determine correlation times in very broad time and temperature ranges. For the studied pore diameters, 2.1--2.9\,nm, we find two crossovers in the temperature-dependent correlation times of liquid water upon cooling. At 220--230\,K, a first kink in the temperature dependence is accompanied by a solidification of a fraction of the confined water, implying that the observed crossover is due to a change from bulk-like to interface-dominated water dynamics, rather than to a liquid-liquid phase transition. Moreover, the results provide evidence that $\alpha$ process-like dynamics is probed above the crossover temperature, whereas $\beta$ process-like dynamics is observed below. At 180--190\,K, we find a second change of the temperature dependence, which resembles that reported for the $\beta$ process of supercooled liquids during the glass transition, suggesting a value of $T_g\!\approx\!185\,$K for interface-affected liquid water. In the high-temperature range, $T\!>\!225\,$K, the temperature dependence of water reorientation is weaker in the smaller C10 pores than in the larger C12 and C14 pores, where it is more bulk-like, indicating a significant effect of the silica confinement on the $\alpha$ process of water in the former 2.1\,nm confinement. By contrast, the temperature dependence of water reorientation is largely independent of the confinement size and described by an Arrhenius law with an activation energy of $E_a\!\approx\!0.5\,$eV in the low-temperature range, $T\!<\!180\,$K, revealing that the confinement size plays a minor role for the $\beta$ process of water.
}
\vspace{1cm}
\end{@twocolumnfalse}
]
\section{Introduction}

Water exhibits many intriguing properties as a bulk and a confined liquid, which are of enormous relevance in nature and technology.\cite{Rupley1991, Rasaiah2008} Bulk water shows several well-known anomalies, including the important density maximum. It was proposed that these anomalies originate from the existence of a second critical point in the deeply supercooled regime, which terminates a liquid-liquid (LL) phase transition between low-density and high-density water phases.\cite{Poole1992,Stanley1998} Experimental validation of the conjectured scenario is, however, not straightforward because bulk water crystallizes in the relevant temperature range, the no-man's land: 150--235\,K. The existence of the LL phase transition thus still remains a subject of controversial scientific discussions.\cite{Ngai2010,Angell2008,Mallamace2012,Bertrand2013}

The properties of confined water can depend on the features of the confining matrix and, hence, deviate from the bulk behavior. It is well known that the melting temperature is reduced until regular crystallization is suppressed when decreasing the size of nanoscopic confinements.\cite{Kittaka2006, YoshidaBellissent2008, Jahnert2008, Findenegg2011, Namba2011} Such suppression of crystallization allows one to keep confined water in the liquid state in the no-man's land and, hence, to explore the possible existence of a LL phase transition. Yet, it remains a crucial question to what extent the nature of water in severe confinements reflects that of water in the bulk liquid. When tackling this issue, it is important to consider that the properties of confined waters can vary across the confining geometry.\cite{Findenegg2011,Soper2012,GalloChen2010} In particular, one may expect that confinement has stronger effects on the behavior of interfacial water near the matrix surface than on that of internal water in the confinement center.  

MCM-41 compounds are ideal matrices to confine water on nanoscales because these silica materials exhibit nanopores with defined and tunable diameters.\cite{Gruen1999} Suppression of water crystallization was reported for MCM-41 with pore diameters $d_p\!\leq\!2.1\,$nm.\cite{YoshidaBellissent2008, Jahnert2008, Findenegg2011}  Previous studies, which exploited this effect to characterize the properties of liquid water in the no-man's land, arrived at different conclusions about the existence of a LL phase transition. On the one hand, neutron scattering (NS) works found a sharp kink in temperature-dependent correlation times of water dynamics at $\sim$225\,K, which was interpreted in terms of a fragile-to-strong  transition, related to a LL phase transition.\cite{YoshidaBellissent2008,Chen2006} On the other hand, dielectric spectroscopy (DS) studies did not observe such transitions, but rather a gentle crossover in the range of 180--190\,K, which was attributed to a subtle interplay of structural $\alpha$ and local $\beta$ relaxations during a glass transition.\cite{Sjostrom2008, Swenson2010, Bruni2011} Our recent $^2$H NMR approach rationalized the observation of two crossovers in temperature-dependent correlation times at 220--230\,K and 180--190\,K, respectively, based on a two-step solidification scenario for water.\cite{Sattig2014} It was proposed that a fraction of water becomes solid at the higher of the two temperatures, leading to a change from bulk-like to interface-dominated dynamics for the other fraction, which stays liquid until it undergoes a confinement-affected glass transition at the lower of the two temperatures, causing another variation of the dynamical behavior.

Here, we investigate the role of the cavity size on water behaviors, in particular, on the proposed relation between crossovers in temperature-dependent correlation times and solidification events. For this purpose, the $^2$H NMR studies on D$_2$O in MCM-41 C10,\cite{Sattig2014} featuring 2.1\,nm confinements, are extended to D$_2$O in MCM-41 materials C12 and C14 with pore diameters up to 2.9\,nm. A combination of spin-lattice relaxation (SLR), line-shape, and stimulated-echo (STE) analyses enables insights into both rates and mechanisms for water reorientation in broad dynamic and temperature ranges, as was also exploited in $^2$H NMR work on water in zeolites.\cite{Pahlke12} In addition, we perform $^1$H SLR measurements for H$_2$O in MCM-41 C12.

\section{Theoretical Background}

While our $^1$H NMR studies are sensitive to the dipolar interaction between the nuclear magnetic moments of the protons, our $^2$H NMR approaches probe the quadrupolar interaction between the nuclear quadrupole moment of the deuterons and an electric field gradient at the nuclear site, as resulting from anisotropic charge distributions within chemical bonds. The associated quadrupolar frequency is given by 
 \begin{equation}
\omega_Q = \pm \frac{\delta}{2} \left[  3 \cos^2 \theta - 1 - \eta\sin^2\theta\cos(2\phi)\right]
\label{eq:quadshift}
\end{equation}
Here, the anisotropy parameter $\delta$ and the asymmetry parameter $\eta$ describe the shape of the electric field gradient tensor and the angles $\theta$ and $\phi$ specify the orientation of this tensor and, hence, of the water molecule, with respect to the external magnetic field $\mathbf{B}_{0}$. The $\pm$ signs correspond to the two allowed transitions between the three Zeeman levels of the deuteron.\cite{SRS1994} For D$_2$O, the anisotropy parameter amounts to $\delta \!=\! 2 \pi \cdot 161\,$kHz, corresponding to a quadrupole coupling constant of $2 \pi \cdot 215$\,kHz. Moreover, the asymmetry parameter is small, $\eta\!=\!0.10$,\cite{Vogel2008, Sattig2014} so that $\omega_Q$ is approximately proportional to the second order Legendre polynomial $P_2(\cos \theta)$, where $\theta$ is the angle between O--D bond and $\mathbf{B}_{0}$ field. Consequently, the fluctuations of the quadrupolar frequency $\omega_Q$ provide access to the rotational correlation function 
\begin{equation}
 F_2 (t) = \frac{\left< P_2 \left[ \cos \theta(0) \right]  P_2 \left[ \cos \theta(t) \right] \right>}{ \left< P_2\left[ \cos \theta(0)  \right]  P_2\left[ \cos \theta(0)  \right] \right>} 
\label{F2}
\end{equation}
of the water molecules. Throughout this contribution, pointed brackets denote the ensemble average.

$^2$H SLR analysis yields information about the spectral density $J_2(\omega)$, which is related to the correlation function $F_2(t)$ via Fourier transformation. For isotropic reorientation, the relaxation time $T_1$ and the spectral density $J_2$ are linked according to\cite{BPP1948} 
\begin{equation}
\frac{1}{T_1} = \frac{2}{15} \, \delta^2 \left[  J_2(\omega_0)  + 4 J_2(2\omega_0) \right] 
\label{eq:bpp}
\end{equation}
with a Larmor frequency $\omega_0$ of $2 \pi \cdot 46.1\,$MHz in our case. When the correlation function of the observed molecular dynamics is a single exponential, $F_2(t) = \exp\left( - t / \tau \right)$, the spectral density takes the Debye form
\begin{equation}
J_D(\omega) = \frac{\tau}{1 + (\omega \tau)^2}
\label{eq:jdebye}
\end{equation}
leading to a $T_1$ minimum for a correlation time $\tau\!\approx\!1/\omega_0\!\approx\!1\,$ns. Thus, upon cooling, the value of $T_1$ decreases  when $\omega_0\tau\!\ll\!1$ at high temperatures, while it increases when $\omega_0\tau\!\gg\!1$ at low temperatures. 

For confined water, the dynamical behavior is more complex and the correlation function is not a single exponential, but distributions of correlation times $G(\log \tau)$ exist. The Cole-Cole (CC) and Cole-Davidson (CD) spectral densities proved useful to describe such complex molecular dynamics:\cite{Beckmann1988}
\begin{equation}
 J_{CC}(\omega) = \frac{ \omega^{-1}\sin\left( \frac{\pi}{2} \beta_{CC} \right)  \left(\omega \tau_{CC} \right)^{\beta_{CC}}    } { 1\! +\! \left(\omega \tau_{CC} \right)^{2\beta_{CC}}\! +\! 2 \cos \left( \frac{\pi}{2} \beta_{CC}\right)   \left(\omega \tau_{CC} \right)^{\beta_{CC}} }
 \label{eq:jcc}
\end{equation}
\begin{equation}
 J_{CD}(\omega) = \frac{\omega^{-1}\sin\left[ \beta_{CD} \arctan (\omega \tau_{CD}) \right]}{\left[ 1+ (\omega \tau_{CD})^2 \right]^{\beta_{CD}/2}}
 \label{eq:jcd}
\end{equation}
They are characterized by the time constants $\tau_{CC}$ and $\tau_{CD}$ and by the width parameters $\beta_{CC}$ and $\beta_{CD}$. The CC and CD spectral densities correspond to symmetric and asymmetric distributions of correlation times $G(\log \tau )$, respectively.\cite{Vogel_EPJ_10} 

When observing the $^2$H NMR line shape of confined liquids, a crossover from a broad Pake spectrum $S_P(\omega)$ to a narrow Lorentzian spectrum $S_L(\omega)$ occurs when isotropic molecular reorientation becomes faster than the time scale of the experiment, $\tau\!\approx \! 1/\delta \! \approx \!1\, \mathrm{\mu s}$, upon increasing the temperature.\cite{Vogel_EPJ_10}  In the solid-echo experiments of the present approach, the transit between the broad and narrow spectra is accompanied by a reduction $R(\log \tau)$ of the signal intensity. Such reduced echo intensity for the intermediate motional regime, $\tau\!\! \approx \!1\, \mathrm{\mu s}$, results because molecular dynamics during the dephasing and rephasing periods of the echo experiment interferes with echo formation.\cite{Vogel2008, Spiess1981, Schmidt1985} 

For broad or bimodal distributions of correlation times, fast ($\tau\!\ll \!1/\delta$) and slow ($\tau \!\gg \!1/\delta$) molecules coexist in a certain temperature range. Then, the observed spectra $S(\omega)$ are approximately described by weighted superpositions of Pake and Lorentzian components:\cite{Roessler1990}
\begin{eqnarray}
S(\omega)&\propto&S_L(\omega)\int_{-\infty}^{\log 1/\delta}G (\log \tau)[1-R(\log \tau)]d\log \tau \nonumber \\
&+& S_P(\omega)\int_{\log 1/\delta}^{\infty}G (\log \tau)[1-R(\log \tau)]d\log \tau
\label{two_phase}
\end{eqnarray}
Here, we take into account that the outcome of the experiment is determined by an effective distribution of correlation times, $G(\log\tau)[1\!-\!R(\log \tau)]$ due to attenuated echo signals of molecules with $\tau\!\approx \! 1/\delta \! \approx \!1\, \mathrm{\mu s}$. Moreover, we use normalized line-shape components $S_L(\omega)$ and $S_P(\omega)$.

$^2$H STE measurements provide access to slow water reorientation,  $10^{-5}\,\mathrm{s}\!<\!\tau\!<\!10^0\,$s. In the STE approach, three pulses divide the experimental time into two short evolution times $t_p\!\ll\!\tau$, which are separated by a mixing time $t_m\!\approx\!\tau$. Using appropriate pulse lengths and pulse phases, it is possible to measure the rotational correlation functions\cite{SRS1994, Fleischer1994, Boehmer2001, Schaefer1995}
\begin{eqnarray}
  F_2^{cc}(t_m,t_p)&\propto&\langle\,\cos\left[\,\omega_Q(0)t_p\right]\cos\left[\, \omega_Q(t_m)t_p\right]\rangle  \label{cc}
  \\
  F_2^{ss}(t_m,t_p)&\propto&\langle\,\sin\left[\,\omega_Q(0)t_p\right]\sin\left[\, \omega_Q(t_m)t_p\right]\rangle
  \label{ss}
\end{eqnarray}
As the angular resolution of STE experiments increases with an extension of the evolution times, measurements of $F_2^{cc}(t_m)$ and $F_2^{ss}(t_m)$ for various values of $t_p$ provides access to the motional mechanism.\cite{Fujara1986, Fleischer1994, Boehmer2001} The angles of elementary rotational jumps can be determined when the time constant of STE decays is analyzed as a function of $t_p$. While the decay time is virtually independent of the value of $t_p$ for large-angle jumps, e.g., tetrahedral jumps, it strongly decreases with increasing evolution time for small-angle jumps, e.g., rotational diffusion. The overall geometry of the reorientation is available from the evolution-time dependence of the residual correlation at long mixing times,\cite{Fujara1986} e.g., from $F_\infty^{cc}(t_p)\!\equiv\!F_2^{cc}(t_m\!\gg\!\tau, t_p)$. In particular, strongly anisotropic reorientation does not cause a complete loss of correlation, but leaves substantial residual correlation. The correlation function $F_2(t_m)$ is obtained from $F_2^{ss}(t_m)$ when the evolution time is short and the anisotropy parameter, as in our case, is negligible so that $\sin(\omega_Q t_p)\approx \omega_Q t_p \propto P_2(\cos \theta)$, see Eq.\ (\ref{F2}). 
 
For the analysis of experimental data, it is important to consider that, in addition to molecular reorientation, spin relaxation damps the STE signal. Therefore, we fit the normalized decays of $F_2^{cc}(t_m)$ and $F_2^{ss}(t_m)$ to the function
\begin{equation}
\left[(1- F_\infty ) \exp \left[ -\left(\frac{t_m}{\tau_K}\right)^{\beta_K}  \right] + F_\infty \right]  \Phi (t_m)\,. 
\label{eq:F2fit}
\end{equation}
Thus, we use a stretched exponential to describe the signal decay due to water dynamics and introduce $F_\infty$ to consider a residual correlation resulting, e.g, in the case of anisotropic reorientation. Moreover, we employ $\Phi(t_m)$ to take into account SLR damping. While the SLR damping of $F_2^{cc}(t_m)$ can be determined in independent SLR measurements, allowing us to fix $\Phi(t_m)$ in the STE analysis, such independent determination of SLR effects is not possible when fitting $F_2^{ss}(t_m)$.\cite{Bohmer1998} Therefore, most of our studies involve the former correlation function.

Owing to broadly diversified dynamics of confined liquids, we use the mean logarithmic correlation time $\tau_m$ to characterize the time scale of water reorientation. In the SLR analysis, $\tau_m\!=\!\tau_{CC}$ for the symmetric CC distribution, while this average value is obtained from
\begin{equation} 
\langle \ln \tau \rangle \equiv \ln \tau_m  =  \ln \tau_{CD}+\psi(\beta_{CD})+\mathrm{Eu} \label{mlntaucd}
\end{equation}
for the asymmetric CD distribution.\cite{Zorn2002} Here, $\mathrm{Eu}\!\approx\! 0.58$ is Euler's constant and $\psi(x)$ denotes the Digamma function. In the STE analysis, the mean logarithmic correlation time can be calculated according to:\cite{Zorn2002}  
\begin{equation}
\langle \ln \tau \rangle \equiv \ln \tau_m = \ln \tau_K + (1-\frac{1}{\beta_K})\mathrm{Eu}\label{mlntaukww}
\end{equation}

\section{Experimental Section}\label{sc:exp}
\subsection{Sample Preparation}\label{prep}
 
The synthesis of the MCM-41 materials was performed according to the standard protocol of Gr{\"u}n et al.,\cite{Grun99} which was adapted by Gr{\"u}nberg and coworkers. \cite{Grunberg2004} Tetraethoxy silane, C$_\mathrm{12}$TAB (dodecyl-trimethyl-ammonium-bromide) or C$_\mathrm{14}$TAB (tetradecyl-trimethyl-ammonium-bromide), distilled water, and 25\% NH$_3$ aq. were used in a ratio of 1 : 0.1349 : 138.8 : 2.8. The template was removed by calcination of the as-synthesized materials at 650\,$^\circ$C for 16 hours resulting in pure mesoporous silica materials. Tetraethoxy silane was acquired from Acros Organics. C$_\mathrm{x}$TABs were purchased from ABCR.  All chemicals were used as received without further purification. Following the nomenclature in the literature, we denote the prepared mesoporous silica as C12 and C14, specifying the C$_\mathrm{x}$TAB chain length. In addition, we use mesoporous silica C10, which was synthesized by Kittaka and coworkers.\cite{YoshidaBellissent2008,Sjostrom2008,Takahara1999,Kittaka2006,Yamaguchi2013} 

\begin{table}[h]
\small
	\caption{Properties of the used MCM-41 materials, as obtained from N$_2$ gas adsorption employing multi-point BET and BJH methods.}
	\label{tbl:samples}
  \begin{tabular*}{0.45\textwidth}{@{\extracolsep{\fill}}l|ccc}
	
    Sample&d$_\mathrm{p}$\,[nm]& V$_\mathrm{p}$\,[cm$^3$/g] &  S$_\mathrm{p}$\,[m$^2$] \\
    \hline
	C14 & 2.93 & 0.86 & 1170   \\  
    C12 & 2.76 & 0.82 & 974 \\ 
	C10\cite{Takahara1999} & 2.14 &  -   & 1096  \\

	\end{tabular*} 
\end{table}

The synthesized MCM-41 samples were characterized employing N$_2$ gas adsorption together with multi-point BET to determine the sample surface area $S_\mathrm{p}$ and the BJH method to obtain the specific pore volume $V_\mathrm{p}$ and the pore diameter $d_\mathrm{p}$.  The properties of the used MCM-41 C12 and C14 materials are shown together with that of MCM-41 C10\cite{YoshidaBellissent2008,Sjostrom2008, Takahara1999,Kittaka2006,Yamaguchi2013,Swenson2013} in Tab.\ \ref{tbl:samples}. 
Prior to use, all MCM-41 matrices were dried by heating to 180\,$^\circ$C for at least 24 hours.

Filling with H$_2$O or D$_2$O employed different routes. The C12 and C14 samples were filled by carefully dripping water onto the dried mesoporous silica materials. The used amounts of water were calculated to obtain fillings of ($100\!\pm\!10$)\% of the pore volumes. For C10, we prepared two samples exhibiting different amounts of excess water outside the pores. Soaking the mesoporous silica in water and subsequent dabbing with filter paper resulted in a sample with a substantial amount of excess water. Afterwards, the material was air-dried, resulting in a sample with a negligible amount of excess water, see below. All samples were sealed in a NMR tube.

\subsection{NMR Experiments}

$^1$H SLR measurements were carried out on a home-built spectrometer working at a Larmor frequency $\omega_0$ of  $2\pi\cdot 360\,$MHz. The 90$^\circ$ pulse length was 5.5\,$\mu$s. The temperature was controlled by pumping gaseous nitrogen, heated from 77\,K to the set temperature, through a custom continuous-flow cryostat, yielding temperature stability of about $\pm$0.5\,K.  

$^2$H NMR experiments were performed using two home-built spectrometers operating at respective Larmor frequencies $\omega_0$ of $2\pi\cdot46.1$\,MHz and $2\pi\cdot46.7$\,MHz. The spectrometers are nearly identical in construction. Both setups utilize a TIC304 MA CryoVac temperature controller together with a Konti CryoVac cryostat, resulting in temperature stability better than $\pm$0.5\,K. In all experiments, the duration of a 90$^\circ$ pulse was in the range 2.1--2.5\,$\mu$s. No dependence of the experimental results on the used $^2$H NMR setup was found. In particular, the difference of the Larmor frequencies $\omega_0$ in both setups is too small to result in resolvable differences of the relaxation times $T_1$.   

$^1$H and $^2$H SLR were studied by observing the recovery of the magnetization, $M(t)$, after saturation. $^2$H NMR spectra were recorded with the solid-echo sequence $90^\circ_x - \Delta - 90^\circ_y$ where the delay $\Delta$ was set to $20\,\mathrm{\mu s}$. In $^2$H STE experiments, we used the pulse sequences $90^\circ_x - t_p - 90^\circ_{-x} - t_m- 90^\circ_x  - t_p$ and  $90^\circ_x - t_p - 45^\circ_y - t_m - 45^\circ_y  - t_p$ to obtain $F_2^{cc}(t_m,t_p)$ and $F_2^{ss}(t_m,t_p)$, respectively. When studying the evolution-time dependence, a fourth 90$^\circ$ pulse was added after a delay of $15\,\mathrm{\mu s}$ to overcome the dead time of the receiver for small values of $t_p$. Appropriate phase cycles were used to cancel out unwanted single-quantum and double-quantum coherences.\cite{Schaefer1995}

\section{Results and Discussion}
\subsection{Spin-Lattice Relaxation}\label{SLR}

\begin{figure}[ht]
	\centering
	\includegraphics[width=7.5cm]{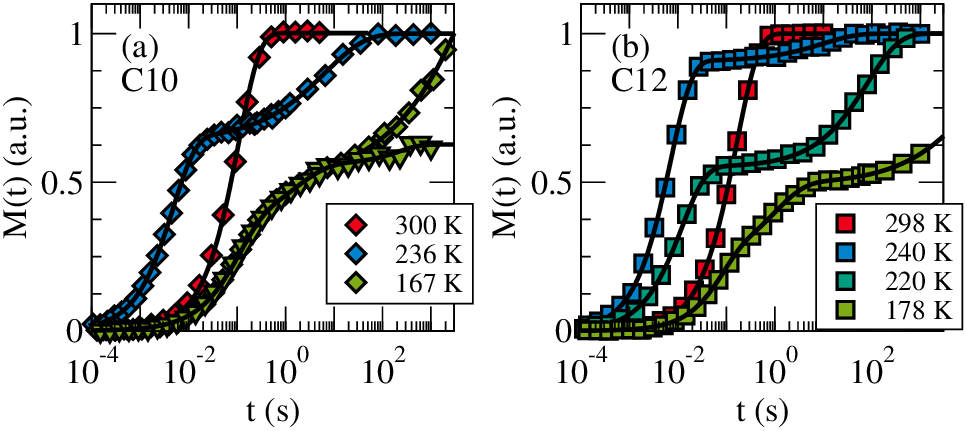}
	\caption{Recovery of $^2$H magnetization after saturation for D$_2$O in (a) C10 and (b) C12. For the C10 matrix, we distinguish samples with substantial (diamonds) and negligible (triangles) amounts of excess water, see Sec.\ \ref{prep}. The solid lines are interpolations with Eq.\ (\ref{eq:fitbu}). }
	\label{Fig_1}
\end{figure}

First, we analyze the buildup of $^2$H magnetization, $M(t)$, to investigate water dynamics in mesoporous silica. Figure \ref{Fig_1} shows results for C10 and C12 samples at several characteristic temperatures. For the studied samples, the buildup of magnetization is monoexponential at ambient temperatures, while it can involve several steps at lower temperatures. To rationalize the relaxation scenario in the latter range, it is useful to recall the freezing behavior of water in the studied MCM-41 materials. While crystallization of excess water outside the pores occurs near 273\,K, freezing of confined water inside the pores depends on the pore size.\cite{Kittaka2006, YoshidaBellissent2008, Jahnert2008, Findenegg2011} In C12, the freezing temperature of water amounts to about 220\,K, whereas regular crystallization of water was found to be largely suppressed in C10. Nevertheless, calorimetric signals were observed in the latter confinement at cryogenic temperatures and attributed to a formation of glassy water or distorted ice.\cite{Namba2010, Kittaka2006, Namba2011, Johari2009}
 
Exploiting this knowledge about the freezing behavior, comparison of buildup curves $M(t)$ for various water contents and confinement sizes enables an assignment of the observed relaxation steps to specific water species. First, we inspect results for C10 samples with substantial and negligible fractions of excess water, see Sec.\ \ref{prep}, in Fig.\ \ref{Fig_1}(a). When excess water exists at 236\,K, a relaxation step at short times, which continues the high-temperature relaxation, is accompanied by a relaxation step at long times, which sets in slightly below 273\,K, but is absent in samples without excess water, as obtained from more detailed analysis, see below. Therefore, we attribute the short-time and long-time steps to confined water and excess water, respectively. As was reported in our previous work,\cite{Sattig2014} a further change of the relaxation behavior occurs near 225\,K, where yet another relaxation starts to split off from the short-time process. To demonstrate the effect, we compare buildup curves for the C10 samples with substantial and negligible fractions of excess water at 167\,K. We observe a third relaxation step at times intermediate between the short-time and long-time relaxations. While this medium-time step, like the short-time step, does not depend on the amount of excess water, the amplitude of the long-time step is again smaller for smaller amounts of excess water. In Fig.\ \ref{Fig_1}(b), we see that the relaxation behavior is qualitatively similar for the C12 sample, in particular, three relaxation steps are distinguishable at sufficiently low temperatures. We conclude that water inside the pores causes not only the short-time relaxation, but also the medium-time relaxation, while ice outside the pores contributes the long-time relaxation. 

Quantitative analysis of the $^2$H SLR behavior enables further insights into the nature of the different water species. Therefore, we fit the buildup curves of the $^2$H magnetization to
\begin{equation}
\frac{M(t)}{M_\infty} = 1 - \sum_{n}{ \alpha_n \exp \left[ - \left( \frac{t}{T_{1,n}}\right)^{\beta_{n}}  \right]  }
 \label{eq:fitbu}
\end{equation}
Here, M$_\infty$ denotes the equilibrium magnetization and the index $n$ refers to the different relaxation steps. For reasons to be specified, we use $n\!=\!l$ (liquid) for the short-time step, $n\!=\!s_i$ (solid inside) for the medium-time step, and $n\!=\!s_o$ (solid outside) for the long-time step. To allow for the possibility of nonexponential $^2$H SLR, we utilize stretched exponentials for the interpolation of the individual contributions. Employing the Gamma function $\Gamma(x)$, the mean SLR times $\langle T_{1,n} \rangle$ are obtained from the fit parameters according to
\begin{equation}
\langle T_{1,n} \rangle=\frac{T_{1,n}}{\beta_n}\;\Gamma\left(\frac{1}{\beta_n}\right)
\end{equation}   

\begin{figure}[ht]
  \centering
  \includegraphics[width=7.0cm]{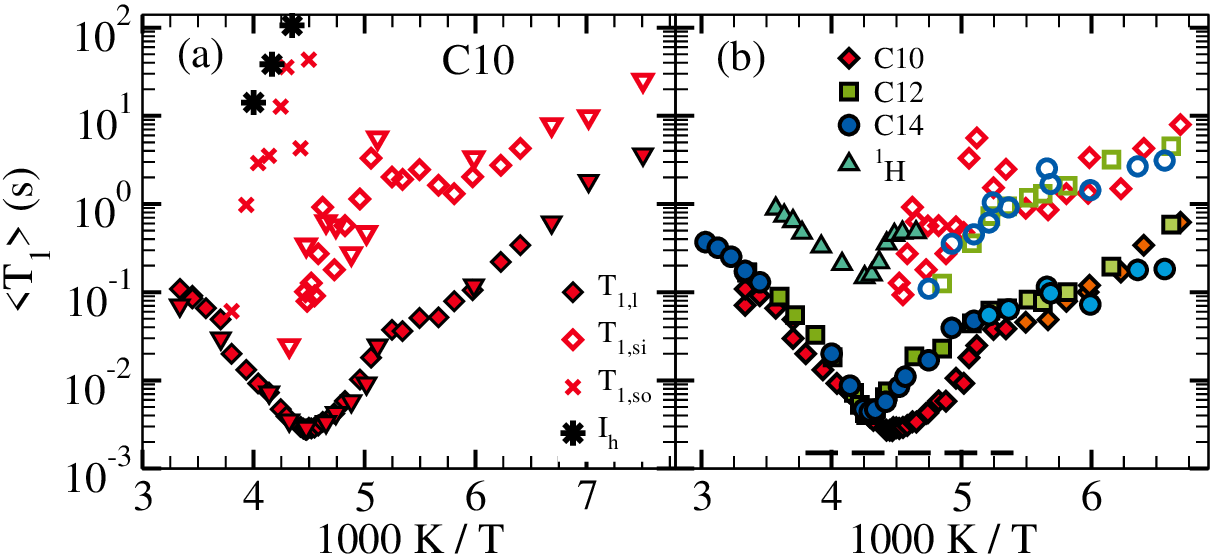}
  \caption{Mean $^2$H spin-lattice relaxation times of the short-time ($\langle T_{1,l} \rangle$, solid symbols), medium-time ($\langle T_{1,s_i} \rangle $, open symbols), and long-time ($\langle T_{1,s_o} \rangle$) steps observed for dynamically distinguishable D$_2$O species in C10, C12, and C14. (a) Results for C10 samples with substantial (diamonds) and negligible (triangles) amounts of excess water are shown, see Sec.\ \ref{prep}.  $\langle T_{1,s_o} \rangle$ is accessible only for a high fraction of excess water and at sufficiently high temperatures. $\langle T_{1}\rangle$ values for bulk ice (I$_h$)\cite{Low} are included. (b) Comparison of the mean relaxation times $\langle T_{1,l} \rangle$ and $\langle T_{1,s_i} \rangle $ for D$_2$O in C10, C12, and C14. The dashed line marks the height of the $^2$H $T_1$ minimum for a Debye process. In addition, the $^1$H spin-lattice relaxation time ($^1$H $T_1$) for H$_2$O inside C12 pores is included.}
  \label{Fig_2}
\end{figure}

Figure \ref{Fig_2} shows results of the SLR analysis for D$_2$O in C10, C12, and C14. In panel (a), we see for C10 that, when a substantial amount of excess water exists and freezes slightly below 273\,K, monomodal SLR turns into bimodal SLR. Upon further cooling, $\langle T_{1,l}\rangle$ decreases, as anticipated for fast dynamics ($\omega_0\tau\!\ll\!1$) in liquid water inside the pores, while $\langle T_{1,s_o}\rangle$ rises, as expected for slow dynamics ($\omega_0\tau\!\gg\!1$) in an ice phase outside the pores. A comparison with $\langle T_{1}\rangle$ values for bulk ice (I$_h$)\cite{Low} supports the latter assignment. Near 225\,K, $\langle T_{1,s_o}\rangle$ becomes too long for a reliable determination in a reasonable amount of time, whereas $\langle T_{1,l}\rangle$ passes a minimum, indicating that confined water exhibits correlation times $\tau\!\approx\!1/\omega_0\!\approx\!1\,$ns. In the same temperature range, the medium-time process appears as a new phenomenon and, hence, a third water species with a distinguishable dynamical behavior emerges. The findings that $\langle T_{1,s_i}\rangle$ ranges well between the other relaxation times and increases when reducing the temperature mean that the dynamics of this water species is intermediate between the slow dynamics of the regular ice and the fast dynamics of the confined liquid. For the C10 sample with a negligible amount of excess water, a splitting into bimodal SLR in the vicinity of 273\,K does not occur, but the medium-time relaxation step again appears near 225\,K. Thereby, the value of $\langle T_{1,s_i}\rangle$ is independent of the amount of excess water, supporting our conjecture that this water species, like the liquid species, resides inside the pores. In panel (b), we see that, qualitatively, both confined water species exhibit comparable SLR behaviors in all studied mesoporous silica. Quantitatively, the $\langle T_{1,l}\rangle$ minimum is shifted to a higher temperature in C12 and C14 as compared to C10 and, on the high-temperature side of the minimum, $\langle T_{1,l}\rangle$ is slightly longer in C12 and C14 than in C10, implying that water dynamics is faster in the larger confinements, at least in the weakly supercooled regime.

Additional knowledge can be obtained from the shape of the relaxation steps, as described by the stretching parameters $\beta_n$. We find that, if existent, the medium and slow relaxations are nonexponential. They are characterized by stretching parameters $\beta_{s_i}\!\approx\!0.6$ and $\beta_{s_o}\!\approx\!0.6$, essentially independent of temperature and sample. By contrast, for all pore sizes, the fast relaxation is exponential above $\sim$185\,K and nonexponential below this temperature. In the latter range, the value of $\beta_l$ continuously decreases when the temperature is decreased, resulting in $\beta_{l}\!\approx\!0.6\!-\!0.7$ at 150-160\,K. In Fig.\ \ref{Fig_2}(b), we use dark and light symbols for $\langle T_{1,l}\rangle$ to indicate this crossover from exponential to nonexponential $^2$H SLR near 185\,K.  

The observations for $\beta_l$ can be rationalized when we consider that water dynamics strongly varies across the pore volume, as found in previous studies on confined water,\cite{Vyalikh07a, Buntkowsky07, Xu14, GalloChen2010, Klameth2013, Klameth2014} including our work on D$_2$O in C10.\cite{Sattig2014} In general, in $^2$H NMR, such distribution of correlation times $\tau$ results in a distribution of relaxation times $T_1$ and, hence, in nonexponential SLR. However, it is crucial for this argument that the correlation time of a molecule does not change during the buildup of the magnetization, which usually occurs on a much longer time scale than the dynamics of the molecule itself. Thus, our finding that the fast relaxation step is exponential above $\sim$185\,K indicates that the correlation times of the associated water fraction are time dependent, implying that these water molecules sample a substantial part of the pore volume on the milliseconds time scale of the buildup process so that the resulting exchange of $\tau$ values averages over any distribution of $T_1$ times, reconstituting exponential relaxation, as expected for a liquid. Vice versa, the continuous development of nonexponentiality found for the fast relaxation step below $\sim$185\,K implies that the molecular diffusion becomes too slow to complete the exploration of a relevant part of the pore volume on the time scale of $\langle T_{1,l}\rangle$, i.e., the system becomes nonergodic. Likewise, the nonexponentiality of the medium-time and long-time steps ($\beta_{s_i},\beta_{s_o}\!<\!1$) shows that the associated fractions of water molecules do not scout different local environments on the time scale of the SLR process and, thus, they are solid. With the same arguments, the existence of several relaxation steps at sufficiently low temperatures provides strong evidence that there are dynamically distinguishable water fractions that do not exchange molecules until the buildup of magnetization is complete. An absence of this exchange is unlikely between two liquid phases. Therefore, our SLR results indicate that liquid and solid water species coexist inside C10 pores below $\sim$225\,K, at least down to $\sim$185\,K, where the fast relaxation step becomes nonexponential, too. 

As for the liquid water fraction, the observed crossover from exponential to nonexponential SLR resembles the situation for supercooled liquids undergoing a glass transition.\cite{Schnauss1990, Boehmer2001} At the glass transition temperature $T_g$, the structural relaxation ceases to ensure ergodicity, resulting in this change of SLR behavior. It is important to notice that, at such temperatures, structural arrangements of supercooled liquids are too slow to be effective for SLR so that the $T_1$ values are determined by faster $\beta$ relaxations rather than by this $\alpha$ relaxation.\cite{Blochowicz} However, changes of $\beta$ dynamics in response to the freezing of $\alpha$ dynamics usually lead to a kink in the temperature-dependent SLR times $\langle T_1(T) \rangle$ at $T_g$.\cite{Schnauss1990, Boehmer2001} For the liquid water fraction in all studied mesoporous silica, such kink of $\langle T_{1,l}(T)\rangle$ is observed at $\sim$185\,K, supporting our conclusion that the structural reorganization of interface-affected water vanishes near 185\,K in a dynamical crossover resembling a glass transition.   

So far, the $^2$H SLR results imply that, in the studied silica matrices, a fraction of confined water solidifies at 220--230\,K, while another fraction of confined water remains liquid below these temperatures. The liquid confined water diffuses in a subvolume of the pores restricted by the silica walls and the solid confined water, at least down to ca.\ 185\,K. The solid water forming at 220--230\,K inside the pores ($s_i$) is different from and may not be mistaken with the frozen water outside the pores ($s_o$) since the latter gives rise to yet another SLR step at even longer times. 

A straightforward relation between relaxation times $T_1$ and correlation times $\tau$ exists for exponential SLR, see Eq.\ (\ref{eq:bpp}). Therefore, we restrict the analysis to the short-time relaxation step at $T\!>\!185$\,K where $\beta_l\!=\!1$ and, consequently, $\langle T_{1,l}\rangle \equiv T_{1,l}$. In Fig.\ \ref{Fig_2}, we see for all studied samples that the minimum value of $T_{1,l}$ is larger than expected for a Debye process, indicating that the correlation function $F_2(t)$ of water reorientation is not a single exponential.\cite{Boehmer2001}  This result is not surprising as dynamical heterogeneities are a characteristic feature of viscous liquids, in particular of confined liquids showing diverse molecular mobilities in various pore regions. To consider the existence of distributions of correlation times $G(\log \tau)$, we utilize the CC and CD spectral densities. Then, the width parameters of these spectral densities, $\beta_{CC}$ and $\beta_{CD}$, can be determined from the minimum values of $T_{1,l}$.\cite{Boehmer2001} The results are compiled in Tab.\ \ref{tbl:paramter}. Assuming temperature-independent width parameters $\beta_{CC}$ and $\beta_{CD}$ and inserting the corresponding spectral densities $J_{CC}(\omega)$ and $J_{CD}$ into Eq.\ (\ref{eq:bpp}), we determine mean logarithmic correlation times $\tau_m$.

\begin{figure}[ht]
  \centering
  \includegraphics[width=7cm]{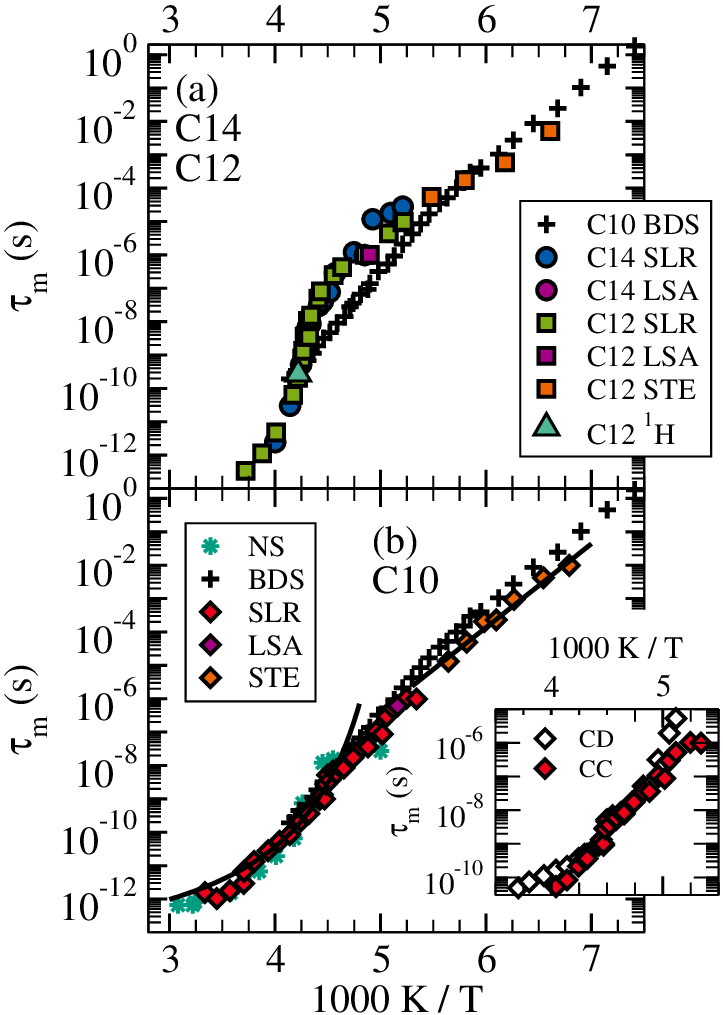}
  \caption{Mean logarithmic correlation times $\tau_m$ of (a) D$_2$O in C12 and C14 and (b) D$_2$O in C10. Data from $^2$H NMR spin-lattice relaxation (SLR), line-shape analysis (LSA), and stimulated-echo experiments (STE) are displayed. For C10, the high-temperature and low-temperature data are fitted to VFT (curved line) and ARR (straight line) laws, respectively. In addition, we show a $\tau_m$ value obtained in $^1$H NMR on H$_2$O in C12 from the $T_1$ minimum. Results from broadband dielectric spectroscopy (BDS)\cite{Sjostrom2008} and neutron scattering (NS)\cite{LiuChen2005} studies on H$_2$O in C10 are included for comparison. The inserted graph compares mean logarithmic correlation times $\tau_m$ resulting for D$_2$O in C10 when the Cole-Cole (CC) and Cole-Davidson (CD) spectral densities are used for spin-lattice relaxation analysis, respectively. The former spectral density was employed to obtain the data in the main graphs. }
 \label{Fig_3}
\end{figure}

In Fig.\ \ref{Fig_3}, we present the results for liquid confined water in C10, C12, and C14. In the inserted graph, mean logarithmic correlations times obtained from the CC and CD spectral densities are compared with each other for the example of C10. For temperatures in the vicinity of the $T_{1,l}$ minimum, both spectral densities yield consistent results. Differences become significant above 240\,K and below 210\,K. Thus, in the temperature range of the proposed LL transition of water, the $^2$H SLR results hardly depend on the choice of the spectral density. Since application of DS to H$_2$O dynamics in MCM-41 pores revealed a CC shape,\cite{Sjostrom2008} we focus on the mean logarithmic correlation times $\tau_m$ resulting from this spectral density in the following. In the main graphs, we see a crossover in the temperature dependence of these time constants at $T\!=\!220$--230\,K, which is prominent for C12 and C14, but exists also for C10.\cite{Sattig2014} At higher temperatures, there are deviations from an Arrhenius (ARR) law, which are strong for the former larger pores and weak for the latter smaller pores. In this range, a Vogel-Fulcher-Tammann (VFT) behavior,
\begin{equation}
\tau_m=\tau_0\exp\left(\frac{B}{T-T_0}\right)\,
\label{VFT}
\end{equation}
may describe the data, as proposed in NS work.\cite{LiuChen2005} At lower temperatures, the temperature dependence of $\tau_m$ is weaker and, at least for C10, consistent with DS data,\cite{Sjostrom2008} which were described by an ARR law. Interestingly, the crossover between VFT-like and ARR-like behaviors for liquid confined water and the emergence of solid confined water occur in the same temperature range 220--230\,K. Therefore, the kink of $\tau_m(T)$ does not necessarily provide evidence for the existence of a LL transition, as will be discussed in more detail below.

\begin{table}[h]
 \small
	\caption{Results of the $^2$H SLR analysis: Minimum SLR time $\langle T_{1,l} \rangle_m$, minimum position $T_m$, width parameter $\beta_{CC}$ of the CC spectral density, and width parameter $\beta_{CD }$ of the CD spectral density.}
	\label{tbl:paramter}
	\begin{tabular*}{0.4\textwidth}{@{\extracolsep{\fill}}c|cccc}		
	\hline
		&  $\langle T_{1,l} \rangle_m$\,[ms] & T$_m$\,[K]& $\beta_{CC}$ & 	$\beta_{CD}$ \\ 
	\hline  
	C14 & 4.34  &  233  & 0.41 & 0.19   \\  
    C12 & 4.04  &  234  & 0.44 & 0.20   \\ 
	C10 & 2.83  &  223  & 0.61 & 0.32   \\
	
	\end{tabular*} 
\end{table}

In the following $^2$H NMR studies, we will exploit the different $^2$H SLR times of the various water species to single out the contribution from the liquid fraction and to suppress that from the solid fractions. In detail, we destroy the magnetization and start the acquisition after delays $\langle T_{1,l}\rangle \!\ll\! t_d \!\ll\! \langle T_{1,{s_i}} \rangle ,\langle T_{1,{s_o}}\rangle$, ensuring that the magnetizations form deuterons in the liquid and solid fractions have recovered to major and minor extents, respectively, when the data are recorded. Hence, such partially relaxed (PR) experiments allow us to ascertain the dynamics of liquid water, essentially unperturbed by contributions from solid water inside or outside the pores. In all experiments, the delay $t_d$ was chosen to maximize the ratio of the signal contributions of the liquid and solid water species.  

Before continuing our $^2$H NMR approach, we analyze $^1$H SLR to ascertain H$_2$O dynamics in C12. We find that, when excess water is absent, the buildup of magnetization is exponential at all studied temperatures in $^1$H NMR, different from the situation in $^2$H NMR. This difference is a consequence of the fact that spin diffusion, i.e., a transfer of magnetization due to flip-flop processes of spins, is faster for $^1$H than for $^2$H mainly due to large differences of the gyromagnetic ratios of these nuclei. Specifically, when solid and liquid fractions are intimately mixed in the pore volumes, spin diffusion is sufficiently fast to exchange magnetization between both species and, hence, to establish a common $T_1$ in $^1$H NMR, while a common $^2$H SLR behavior is not necessarily established, as is observed in our measurements. In Fig.\ \ref{Fig_2}, we see that $^1$H $T_1$ exhibits a clear minimum at $\sim$240\,K, indicating a correlation time of $\tau\!\approx\!1/\omega_0\!=\!0.44$\,ns at this temperature. Inspection of Fig.\ \ref{Fig_3} reveals that this correlation time from $^1$H SLR well agrees with that from $^2$H SLR. Hence, differences between H$_2$O and D$_2$O dynamics in C12 are small. 

\subsection{Line-Shape Analysis}

\begin{figure}[ht]
 \centering
 \includegraphics[width=7.0cm]{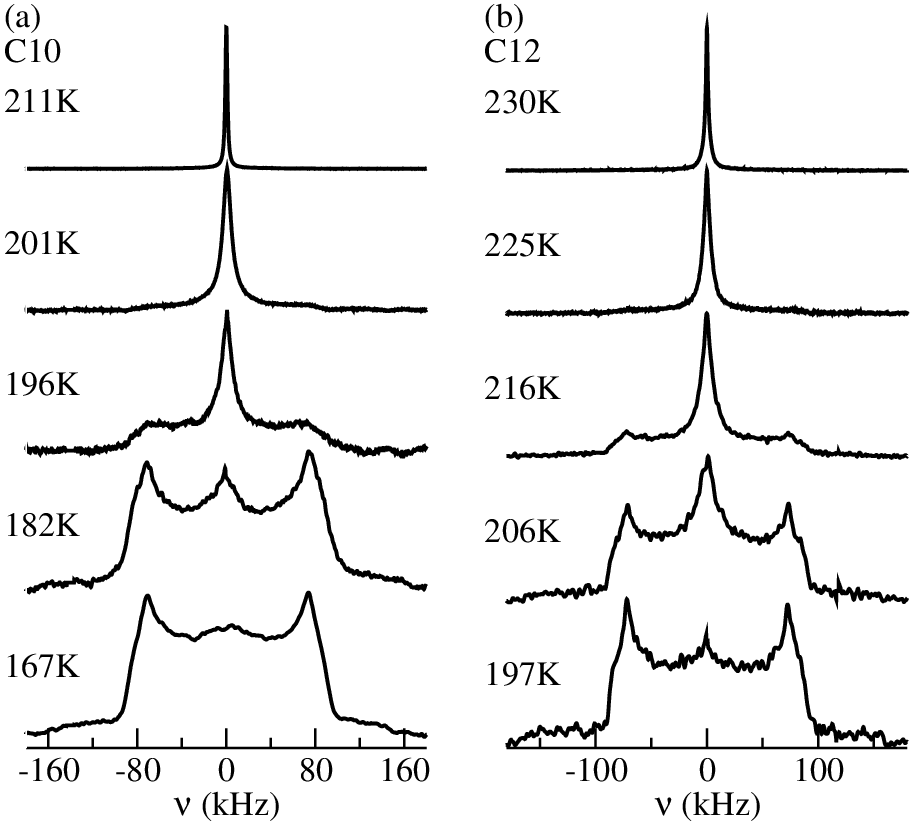}
  \caption{Solid-echo spectra obtained from partially relaxed experiments at the indicated temperatures: (a) D$_2$O in C10 and (b) D$_2$O in C12.}
  \label{Fig_4}
\end{figure}

Figure \ref{Fig_4} shows $^2$H NMR spectra of D$_2$O in C10 and C12 from PR experiments. While a narrow Lorentzian line is observed at sufficiently high temperatures, indicating that the rotational motion of water is fast ($\tau\!\ll \!1\, \mathrm{\mu s}$) and isotropic, a broad Pake spectrum is found at sufficiently low temperatures, revealing that the reorientation dynamics of water is slow ($\tau\!\gg \!1\, \mathrm{\mu s}$). In the intermediate temperature range, the spectra can be described as a weighted superposition of Lorentzian and Pake components, suggesting that fast ($\tau\!\ll \!1\, \mathrm{\mu s}$) and slow ($\tau\!\gg \!1\, \mathrm{\mu s}$) molecules coexist. Comparing the spectra for the different pore sizes, we observe that the crossover between the Lorentzian and Pake shapes occurs at lower temperatures for C10 than for C12, indicating that water dynamics is faster in the smaller than in the larger confinement in the vicinity of 200\,K. 

On first glance, it may be tempting to assign the Lorentzian and Pake lines to the liquid and solid water species, respectively. However, one has to bear in mind that PR experiments are designed to suppress the latter contribution. In previous work,\cite{Sattig2014} it was confirmed that a Pake contribution resulting from solid water is eliminated in the PR spectrum of D$_2$O in C10. Therefore, the assignment of the Lorentzian and Pake components to the liquid and solid fractions of confined water was ruled out. Rather, it was argued that the liquid water by itself gives rise to a superposition of narrow and broad lines. Specifically, a coexistence of Lorentzian and Pake contributions can result from not only bimodal dynamics, but also a broad distribution $G(\log \tau)$, see Eq.\ (\ref{two_phase}), as found for the liquid water species in our SLR analysis. Then, fast ($\tau\!\ll \!1\, \mathrm{\mu s}$) and slow ($\tau\!\gg \!1\, \mathrm{\mu s}$) molecules from the distribution are at the origin of the narrow and broad lines, respectively, while molecules with correlation times $\tau\!\approx \!1\, \mathrm{\mu s}$ can be neglected because their signals are poorly refocused during the used solid-echo pulse sequence, as described by the reduction factor $R(\log \tau)$, see Eq.\ (\ref{two_phase}). In this scenario, the weighting factor of the Lorentzian line, $W(T)$, i.e., the contribution of this line-shape component to the total spectral intensity, is expected to continuously decrease upon cooling as the distribution $G(\log \tau)$ shifts to longer times and, hence, the fraction of fast molecules decreases. Such scenario was also proposed for benzene,\cite{Gedat02} isobutyric acid,\cite{Vyalikh07b} bipyridine,\cite{DeSousa12} and naphtalene\cite{Grunberg13} confined in mesoporous silica. By contrast, in $^2$H NMR on D$_2$O in zeolites,\cite{Pahlke12} the observation of a comparable coexistence of narrow and broad line-shape features was not attributed to a broad distribution $G(\log \tau)$, but to a specific motional mechanism, which involves tetrahedral jumps and $\pi$ flips with a single correlation time $\tau$.   

\begin{figure}[ht]
 \centering
 \includegraphics[width=7.0cm]{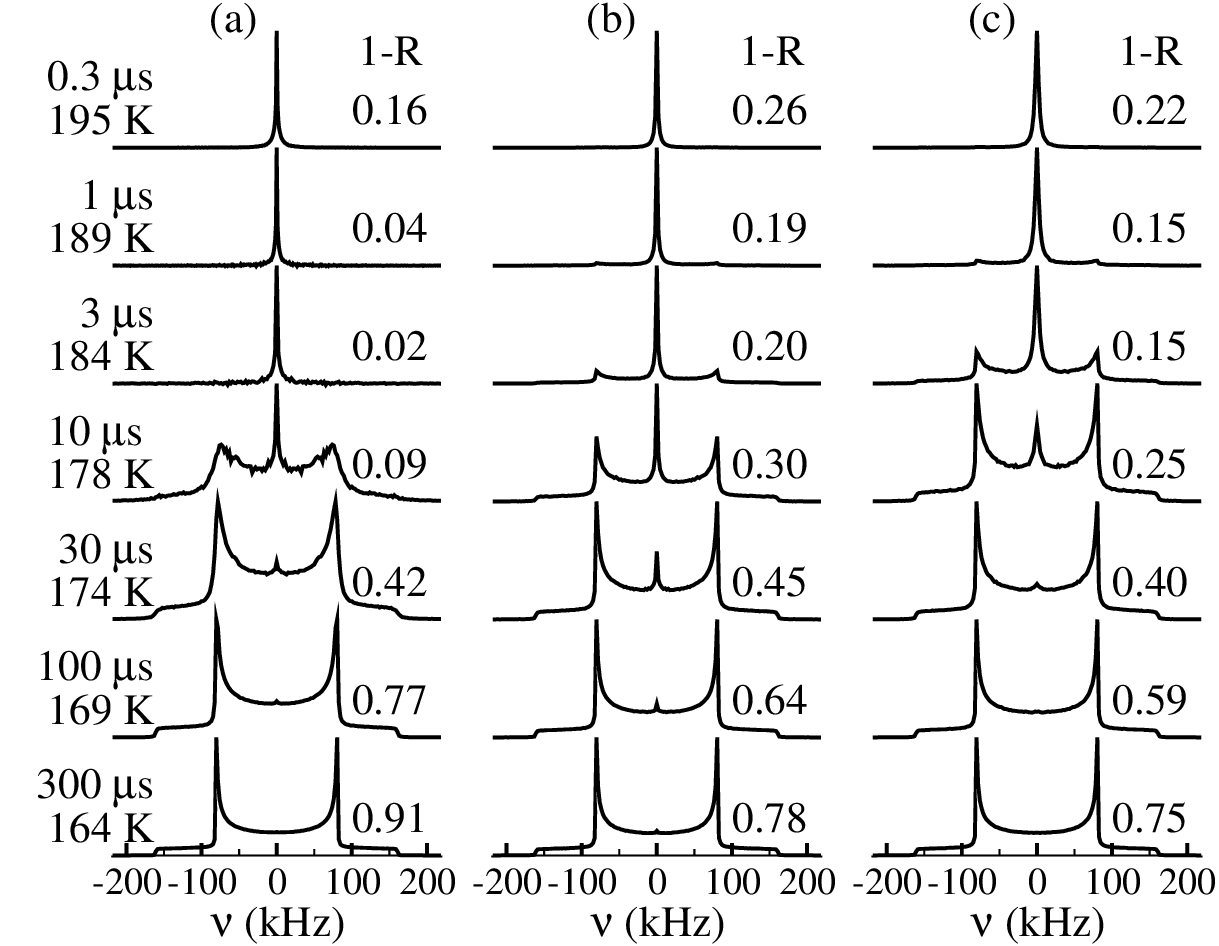}
  \caption{Solid-echo spectra obtained from random-walk simulations for (a) tetrahedral jumps with a single correlation time $\tau$, (b) tetrahedral jumps with a distribution $G(\log \tau)$, and (c) isotropic random jumps with a distribution $G(\log \tau)$. For the simulations, $\delta\!=\!2\pi\cdot161\,$kHz and $\eta\!=\!0$ were assumed and the solid-echo delay was set to the experimental value $\Delta\!=\!20\,\mu$s. Logarithmic Gaussian distributions $G(\log \tau)$ with a width parameter of $\sigma\!=\!2.1$ were used, which closely resemble the CC distribution for C10 from the present $^2$H SLR analysis. Specifically, both distributions have the same second moment. The characteristic time constants of the reorientation processes and the corresponding temperatures, as obtained for C10, see Fig.\ \ref{Fig_3}, are indicated on the left hand side. More precisely, the jump correlation times $\tau_j$ characterizing the exponential distribution of waiting times between consecutive reorientation events are indicated. For isotropic random jumps, the jump correlation time $\tau_j$ can be identified with the rotational correlation time $\tau_2$ of $F_2$, while $\tau_j=\frac{4}{3}\tau_2$ for tetrahedral jumps, see Fig.\ \ref{Fig_9}. For the distributions, the indicated values are mean logarithmic time constants. In addition, the reduced solid-echo intensities $1\!-\!R$ are specified for the various dynamical scenarios.}
  \label{Fig_5}
\end{figure}

To ascertain the role of dynamical heterogeneities for the line shape, we calculate solid-echo spectra from random-walk simulations. The methodology of these calculations was described in previous works.\cite{Vogel2000, Vogel2001b} In Fig.\ \ref{Fig_5}(a) and (b), we show simulated spectra for tetrahedral jumps with single and distributed correlation times. In both cases, narrow and broad spectral patterns can coexist during the line-shape transition. However, such spectral pattern is found in a narrower dynamic range and, hence, a narrower temperature range for single than for distributed time constants. When using the above findings to map correlation times onto experimental temperatures, we observe that the coexistence range from the simulations for a single correlation time does not agree with that found in the measurements, while the expected and observed Lorentzian-Pake superposition regimes between about 175 and 200\,K are in reasonable agreement for a Gaussian distribution $G(\log \tau)$ with $\sigma\!=\!2.1$, which resembles the CC distribution from the above SLR data. Thus, line-shape analysis provides evidence that confined water exhibits pronounced dynamical heterogeneities. On the other hand, this approach does not allow us to distinguish between tetrahedral jumps and isotropic reorientation since both mechanisms lead to comparable spectra when the same distribution $G(\log \tau)$ is used in the simulations, see Fig.\ \ref{Fig_5}(b) and (c). 

\begin{figure}[ht]
  \centering
  \includegraphics[width=6cm]{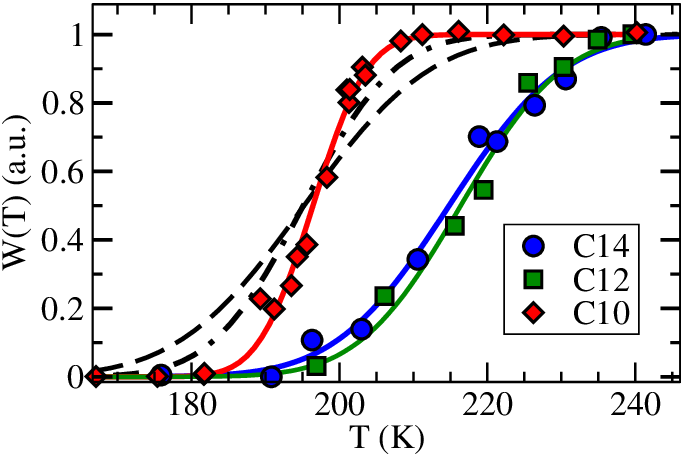}
  \caption{Weighting factor $W(T)$ describing the contribution of the Lorentzian line to the total intensity of partially relaxed spectra for D$_2$O in C10, C12, and C14. The solid lines are guides to the eye. The other lines are predictions for C10 calculated from distributions of correlation times reported in a dielectric spectroscopy study on H$_2$O in C10.\cite{Sjostrom2008} The dashed and dotted results were obtained when neglecting and considering the reduction factor $R(\log \tau)$ inherent to solid-echo experiments, respectively. See text for details.}
  \label{Fig_6}
\end{figure}

To further explore the conjecture that a broad distribution $G(\log \tau)$ governs the line shape of liquid confined water, we determine the weighting factor $W(T)$ from Lorentzian fits of the central region of PR spectra. In Fig.\ \ref{Fig_6}, a continuous temperature evolution of $W(T)$ is obvious for D$_2$O in C10, C12, and C14. A similar increase of the weighting factor with increasing temperature is observed for C12 and C14, while the rise of $W(T)$ occurs at lower temperatures and in a narrower range for C10. These observations are consistent with the outcome of our SLR analysis. More precisely, when we use the criterion $W(T)\!=\!0.5$ to extract a correlation time $\tau_m\!=\!1\,\mu$s, we find good agreement with the time constants from SLR analyses, see Fig.\ \ref{Fig_3}. Moreover, higher values of the width parameters $\beta_{CC}$ and $\beta_{CD}$ for C10 than for C12 and C14, see Tab.\ \ref{tbl:paramter}, imply a narrower distribution $G(\log \tau)$ for the former confinement, in harmony with a narrower temperature range of Lorentzian-Pake coexistence. 

For a final check of the hypothesis that dynamical heterogeneities associated with the liquid fraction are responsible for the line shape, we calculate expectations for the Lorentzian contribution using the distribution $G(\log \tau)$ reported in a DS study on H$_2$O in C10.\cite{Sjostrom2008} First, we neglect the reduction factor $R(\log \tau)$ associated with solid-echo experiments and obtain the Lorentzian contribution directly from $W(T)\! =\! \int^{-6}_{-\infty} G(\log \tau)\,d \log \tau$, see Eq.\ (\ref{two_phase}). In Fig.\ \ref{Fig_6}, it is evident that the calculated and measured weighting factors roughly agree, but the transition range is wider for the former than for the latter. Therefore, it is necessary to consider the reduction factor of the experiment and to employ an effective distribution of correlation times for the calculations, explicitly, $W(T)\!= \!\int^{-6}_{-\infty} G(\log \tau ) [1\!-\!R(\log \tau))] \,d \log \tau$, see Eq.\ (\ref{two_phase}). The exact shape of $R(\log \tau)$ depends on the mechanism for molecular dynamics, see Fig.\ \ref{Fig_5}(b) and (c). For isotropic random jumps, the full width at half maximum was found to amount to about two orders of magnitude.\cite{Vogel2008} Therefore, we use a logarithmic Gaussian distribution with a width parameter $\sigma \!=\! 0.85$, which is centered at $\log (\tau/s) \!= -6$. In Fig.\ \ref{Fig_6}, we see that this approach reasonably well describes the experimental data. Hence, we can conclude that the liquid fraction of confined water exhibits pronounced dynamical heterogeneities, leading to a superposition of Lorentzian and Pake lines in the NMR spectra at intermediate temperatures.

\subsection{Stimulated-Echo Experiments}

\begin{figure}[ht]
  \centering
  \includegraphics[width=7.0cm]{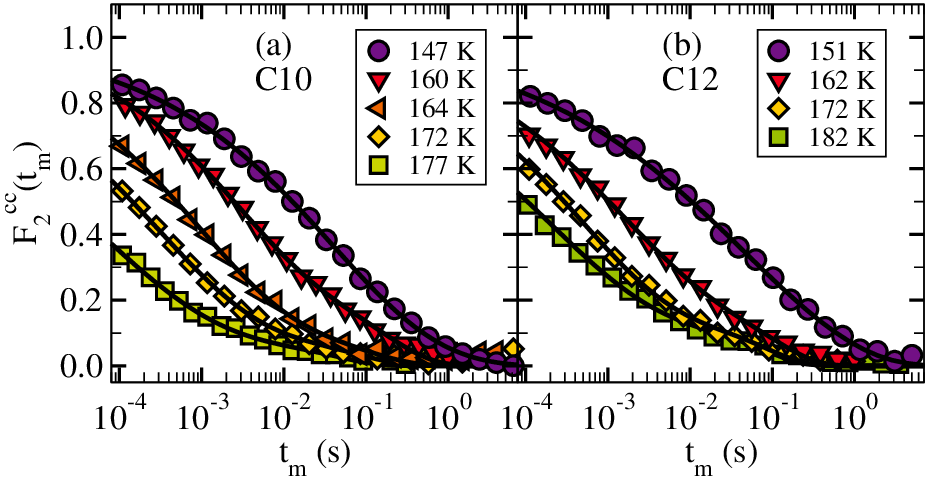}
  \caption{Rotational correlation functions $F_2^{cc}(t_m)$ of (a) D$_2$O in C10 and (b) D$_2$O in C12. The data were obtained from partially relaxed experiments for an evolution time of $t_p\!=\!9\,\mu$s. The lines are interpolations with Eq.\ (\ref{eq:F2fit}). In these fits, we exploit that the spin-lattice relaxation damping $\Phi(t_m)$ is known from independent analysis in Sec.\ \ref{SLR}. We obtain stretching parameters in the range $\beta_K\!=\!0.25\!-\!0.39$ for the studied samples and temperatures. }
  \label{Fig_7}
\end{figure}

Finally, we use $^2$H STE experiments to study slow water dynamics at low temperatures. Again, we perform PR measurements to focus on the behavior of liquid confined water. Figure \ref{Fig_7} shows $F_2^{cc}(t_m)$ for D$_2$O in C10 and C12 at various temperatures. For both samples, the decays are not exponential and shift to longer times when the temperature is decreased, in harmony with water dynamics that is governed by strong heterogeneity and slows down upon cooling. At given temperatures $T\!\leq\!185\,$K, the loss of correlation is comparable for  the C10 and C12 samples, while relaxation-time and line-shape analyses indicated that water dynamics is faster in smaller pores at higher temperatures. Furthermore, the STE decays exhibit a weak two-step character with a short-time decay due to molecular dynamics and a long-time decay due to spin relaxation. Specifically, good interpolations of the experimental data result when fitting $F_2^{cc}(t_m)$ to Eq.\ (\ref{eq:F2fit}) and fixing the parameters of the SLR damping $\Phi(t_m)$ at the values obtained from the SLR analysis in Sec.\ \ref{SLR}.

For a characterization of the decay owing to water reorientation, we calculate mean logarithmic correlation times $\tau_\mathrm{m}$ using the fit results for $\tau_K$ and $\beta_K$ together with Eq.\ (\ref{mlntaukww}). The results are included in Fig.\ \ref{Fig_3}. It is evident that the time constants hardly depend on the pore size. For D$_2$O in both C10 and C12, an ARR law with an activation energy of $E_a\!=\!0.5$\,eV describes the STE data. Furthermore, we see that the temperature dependence resulting from the STE analysis at lower temperatures $T\!\leq\!185\,$K is somewhat weaker than that obtained from the SLR data at higher temperatures. In both temperature ranges, the present NMR data are fully consistent with prior DS results for H$_2$O in C10,\cite{Sjostrom2008, Swenson2010, Bruni2011} which reported a mild change of the temperature dependence at 180--190\,K. 

%Hence, our approach unravels that . A crossover at 220--230\,K occurs in the temperature range of a kink in the temperature dependence reported in NS works,\cite{YoshidaBellissent2008,Chen2006} while a crossover at 180--190\,K is in harmony with DS results.\cite{Sjostrom2008, Swenson2010, Bruni2011}

\begin{figure}[ht]
  \centering
  \includegraphics[width=7.0cm]{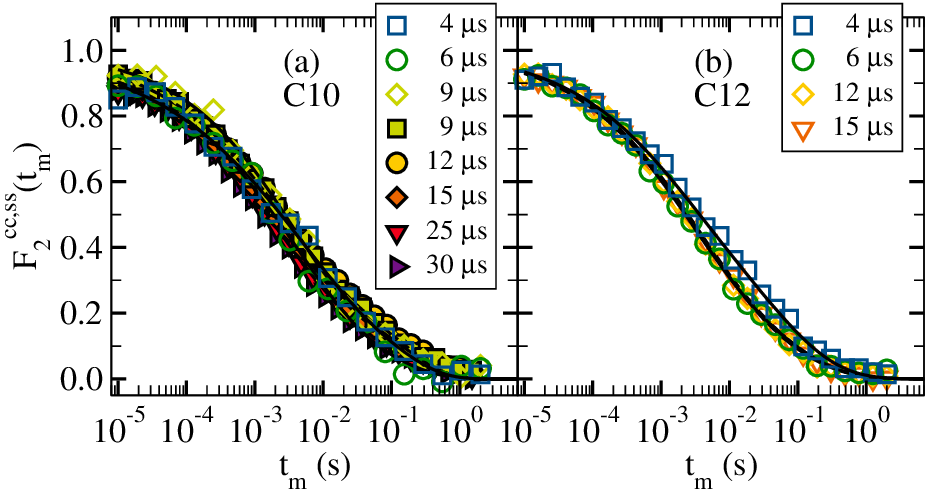}
  \caption{Rotational correlation functions $F_2^{cc}(t_m)$ (solid symbols) and $F_2^{ss}(t_m)$ (open symbols) for various values of the evolution time $t_p$ at 160\,K: (a) D$_2$O in C10  and (b) D$_2$O in C12. All data were obtained from partially relaxed experiments. The lines are interpolations with Eq.\ (\ref{eq:F2fit}).}
  \label{Fig_8}
\end{figure}

\begin{figure}[ht]
  \centering
  \includegraphics[width=7.0cm]{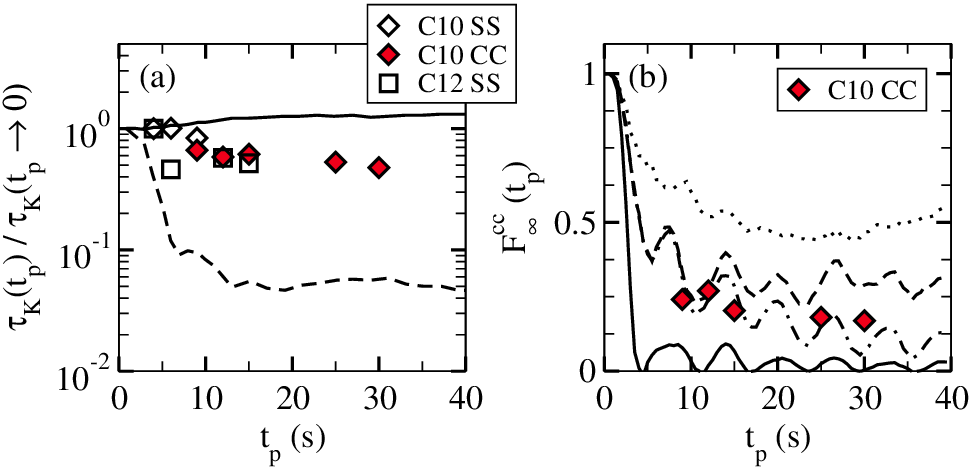}
  \caption{Parameters characterizing the dependence of the correlation functions $F_2^{cc}(t_m,t_p)$ and $F_2^{ss}(t_m,t_p)$ on the length of the evolution time: (a) time constants $\tau_K^{cc}(t_p)$ and $\tau_K^{ss}(t_p)$ and (b) residual correlation $F_\infty^{cc}(t_p)$. The results were obtained by fitting stimulated-echo decays of D$_2$O in C10 and C12 at $T\!=\!160\,$K. In panel (a), the experimental data $\tau_K^{cc,ss}(t_p)/\tau_K^{ss}(t_p\!\rightarrow\!0)$ are compared with simulation results for tetrahedral jumps (solid line) and isotropic $10^\circ$ jumps (dashed line). In panel (b), the measured data are contrasted with calculated expectations for isotropic reorientation (solid line), $180^\circ$ jumps about the molecular symmetry axis (dotted line), and tetrahedral jumps. In the latter case, we distinguish between exact tetrahedral jumps (dashed line) and distorted tetrahedral jumps (dash-dotted line), where we randomly choose orientations on the edges of cones with semi-opening angles of $3^\circ$ around the tetrahedral positions.}
  \label{Fig_9}
\end{figure}

To obtain insights into the mechanism for water reorientation at low temperatures $T\!<\!185\,$K, we exploit that the angular resolution of STE experiments is enhanced when the evolution time is extended. In Fig.\ \ref{Fig_8}, we show correlation functions $F_2^{cc}(t_m)$ and $F_2^{ss}(t_m)$ at 160\,K for various lengths of the evolution time $t_p$. It can be seen for D$_2$O in C10 and C12 that the dependence of the correlation functions on the evolution time is weak. For a quantitative analysis, the STE decays are again interpolated with Eq.\ (\ref{eq:F2fit}). 

Figure \ref{Fig_9} shows the resulting time constants and residual correlations. In Fig.\ \ref{Fig_9}(a), we see that the time constants $\tau_K^{cc}(t_p)$ and $\tau_K^{ss}(t_p)$ are basically independent from the value of the evolution time and the type of the correlation function. This independence indicates that overall water reorientation results from large-angle rather than small-angle elementary jumps, as is confirmed by comparison with curves calculated for different motional models.\cite{Vogel2008} Specifically, the experimental findings are in reasonable agreement with expectations for jumps about the tetrahedral angle, while they are in contrast to simulations for isotropic jumps about an angle of $10^\circ$. In passing, we note that the similarity of the $F_2^{cc}(t_m)$ and $F_2^{ss}(t_m)$ data and the weak evolution-time dependence of the results indicate that the choice of the experimental parameters is not critical when we use STE experiments to determine the temperature dependence of water dynamics, as was tacitly assumed in the above analysis. 

As for the residual correlation, determination of reliable values is possible for $F_\infty^{cc}$, but not for $F_\infty^{ss}$ since the long-time plateau of the correlation functions is affected by relaxation effects, which are known from independent experiments in the case of $F_2^{cc}$, but not in the case of $F_2^{ss}$.\cite{Bohmer1998} In Fig.\ \ref{Fig_9}(b), residual correlations $F_\infty^{cc}(t_p) \approx 0.2 \pm 0.1$ can be observed for the studied evolution times with a mild tendency to decrease when this time interval is extended. Thus, the fit results confirm the above conclusion from visual inspection that the probed low-temperature dynamics of liquid water does not cause complete decays of the correlation functions. Further information is available when we contrast the experimental data with expectations for isotropic reorientation, tetrahedral jumps, and $180^\circ$ jumps around the molecular symmetry axis. The observed residual correlation is significantly higher and lower than the expectations for isotropic reorientation and $180^\circ$ jumps, respectively. Hence, water reorientation does exhibit significant anisotropy at the studied temperatures, but it is not given by a simple two-site jump related to the molecular symmetry. Rather, the observations for $F_\infty^{cc}(t_p)$ are consistent with a tetrahedral jump, in particular, when we allow for some distortions ($\pm3^\circ$) of the molecular orientations, as expected in disordered hydrogen-bond networks. However, a unique determination of the motional geometry is not possible since similar behaviors $F_\infty^{cc}(t_p)$ are produced by other anisotropic motions \cite{Fleischer1994} and uncertainties in the determination of the plateau values result from imperfections in the suppression of the water species with the longer relaxation time $T_{1,s_i}$ in the partially relaxed measurements. 

\section{Conclusion}

We employed a combination of $^1$H and $^2$H NMR methods to investigate H$_2$O and D$_2$O in MCM-41 C10, C12, and C14. For the studied matrices, we found that all confined water is in the liquid state above 220--230\,K, whereas fractions of liquid water and solid water coexist inside the pores below this temperature range, as indicated by an observation of separate SLR steps. Most probably, the liquid and solid fractions can be identified with interfacial and internal waters, respectively. 

For liquid confined water, the temperature-dependent rotational correlation times show two crossovers. A crossover at 220--230\,K accompanies the appearance of solid confined water. While a VFT behavior roughly describes water reorientation in the confined liquid above the crossover, clear deviations exist below. A crossover at 180--190\,K is indicated by a mild weakening of the temperature dependence, which is better seen in prior DS results\cite{Sjostrom2008, Swenson2010, Bruni2011} than in the present NMR data. The latter effect comes along with a change of the SLR behavior. While an exponentiality of SLR together with an observation of Lorentzian line shapes revealed that liquid water molecules show an isotropic reorientation associated with an exploration of a substantial part of the pore volume above the latter crossover, a nonexponentiality of SLR and an existence of finite residual correlation provided evidence that these molecules exhibit an anisotropic and localized reorientation below. These observations imply that the reorganization of the water network ceases at 180--190\,K.

Based on these results, we propose the following scenario. Above 220--230\,K, our NMR studies probe the $\alpha$ process of water molecules in all pore regions. In this temperature range, the temperature dependence becomes stronger and, hence, the bulk behavior is approached when increasing the pore radius, as expected. Two findings show that an observation of a kink in temperature-dependent correlation times of confined water at $\sim$225\,K is not sufficient to conclude on the existence of a LL phase transition at this temperature. First, an appearance of solid water, most likely, freezing or vitrification\cite{Swenson2010} of water in the inner area of the pore, accompanies the alteration of the temperature dependence found for liquid water at 220--230\,K, implying that the space available to the liquid fraction changes from 3D-like to 2D-like, i.e., to the outer area of the pore. Second, liquid water exhibits a prominent $\beta$ process below 220--230\,K, implying that a splitting of the $\beta$ process from the $\alpha$ process contributes to changes in temperature-dependent correlation times in this range. Notwithstanding the foregoing, the present results do not allow us to rule out a LL phase transition for \emph{bulk} water.

In the temperature range from $\sim$225\,K to $\sim$185\,K, the NMR results related to liquid water start to be governed by the $\beta$ process, but these water molecules still explore a substantial part of the pore volume on a longer time scale, restoring ergodicity and, thereby, exponentiality of SLR. Thus, structural relaxation still occurs. While the strong confinement provided by the pore walls and water nanosolids may alter the nature of the $\alpha$ process, the scenario, in a sense, resembles that in bulk supercooled liquids exhibiting $\alpha$-$\beta$ splitting upon cooling. Compared to other liquids,\cite{Vogel2001a, Vogel2001b, Doss2002} the $\beta$ process of water is, however, related to molecular reorientation with much larger amplitude and, hence, it dominates the loss of orientational correlation, rendering an observation of the $\alpha$ process of water difficult. Because of this difficulty, there is still some room for discussions about the existence of the $\alpha$ process of confined and bulk waters below 220--230\,K.\cite{Swenson2010}     

The observations for liquid water at 180--190\,K resemble changes reported for the $\beta$ process of supercooled liquids in response to a glass transition. In particular, the SLR of liquid water changes from exponential to nonexponential, implying that the $\alpha$ process ceases to restore ergodicity in this temperature range, as observed for bulk liquids at $T_g$.\cite{Schnauss1990, Boehmer2001} Also, the temperature dependence of the SLR times of the liquid fraction, $\langle T_{1,l}\rangle$, becomes weaker. For bulk liquids, such effect results from changes of the $\beta$ process owing to the freezing of the $\alpha$ process during vitrification.\cite{Boehmer2001} Therefore, we propose that interfacial water exhibits a glass transition at 180--190\,K. This conjecture is supported by the observation of calorimetric signals for confined water at such temperatures.\cite{Yamaguchi2013, Namba2011} Due to the strong geometrical restrictions, it is, however, questionable whether the present and previous results for confined water are of any relevance for the glass transition of bulk water. 

Below 180--190\,K, the $\beta$ process of water is associated with anisotropic and localized motion, yet it continues to involve reorientations about large angles, in agreement with findings for water dynamics at other interfaces.\cite{Vogel2008, Lusceac2010a, Lusceac2010b, Lusceac2011b}  At these lower temperatures, the correlation times of water reorientation follow an ARR law with an activation energy of $E_a\!\approx\!0.5$\,eV. They hardly depend on the pore size, at variance with our observations at higher temperatures, providing further evidence that distinct relaxation processes are probed in different temperature regimes. 

There is a certain similarity between the present findings for water dynamics below $\sim$225\,K and previous results for small mobile molecules close to large immobile molecules in binary mixtures, which were interpreted in terms of a type A glass transition,\cite{Blochowicz2012} as predicted within mode-coupling theory.\cite{Bosse1996} For such type A glass transition, a nonergodicity level continuously develops upon cooling, in contrast to a discontinuous jump of the nonergodicity parameter at the critical temperature $T_c$ for a type B glass transition. Thus, in future studies, the temperature evolution of the long-time plateau of the correlation function should be ascertained in more detail to distinguish between possible descriptions of water dynamics in the low-temperature regime.     

Altogether, the present $^2$H NMR studies provide evidence that, upon cooling, the dynamical behavior of liquid water in silica pores with diameters of 2--3\,nm changes at 220--230\,K and 180--190\,K. These changes are closely related to a formation of crystalline or glassy water species in combination with an onset of a pronounced $\beta$ relaxation, which exhibits a number of common features for water in various confinements and mixtures.

\section*{Acknowledgments}
We thank Jan Swenson (Chalmers University) for providing us with MCM-41 C10 and the Deutsche Forschungsgemeinschaft for funding through Grants No.\ Bu-911/18-1, Fu-308/16-1, and Vo-905/9-1.

\balance
\footnotesize{
\bibliography{Biblio_PCCP} %your .bib file
\bibliographystyle{rsc} %the RSC's .bst file
}

\end{document}